\documentclass[epjST]{svjour}
\usepackage{verbatim,csquotes,bbm,mathrsfs,amsmath,amssymb}
\usepackage{graphicx,color,mathrsfs}
\graphicspath{{./figures/}}
\usepackage{tikz}
\usepackage{circuitikz}
\usepackage{ulem}

\usepackage[psdextra]{hyperref} 

\usepackage[backend=bibtex,sorting=none,style=numeric-comp,url=false,giveninits=true]{biblatex}
\DeclareFieldFormat{doi}{\mkbibacro{DOI}\addcolon\space\url{https://doi.org/#1}}
\newcommand{\bra}[1]{\langle {#1} \vert}
\newcommand{\ket}[1]{\vert {#1}\rangle}
\newcommand{\braket}[2]{\langle{#1} \vert {#2}\rangle}
\newcommand{\ketbra}[2]{\vert {#1} \rangle \! \langle{#2}\vert}

\usepackage{graphicx}
\begin{document}

\title{Integrated conversion and photodetection of virtual photons in an ultrastrongly coupled superconducting quantum circuit.}
\author{Luigi Giannelli\inst{1,2,3}\fnmsep\thanks{\email{luigi.giannelli@dfa.unict.it}} \and Giorgio Anfuso\inst{2} \and Miroslav Grajcar\inst{4} \and Gheorghe Sorin Paraoanu\inst{5} \and
  Elisabetta Paladino\inst{1,2,3} \and  Giuseppe Falci\inst{1,2,3}}
\institute{CNR-IMM, UoS Universit\`a, 95123, Catania, Italy \and Dipartimento di
  Fisica e Astronomia ``Ettore Majorana'',  Universit\`a di Catania, Via S.
  Sofia 64, 95123, Catania, Italy \and INFN Sez. Catania, 95123  Catania, Italy
  \and Department of Experimental Physics, Comenius University, SK-84248 Bratislava, Slovakia \and QTF Centre of Excellence, Department of Applied Physics,
  Aalto University, P.O. Box 15100, FI-00076 AALTO, Finland}
\abstract{The ground-state of an artificial atom ultrastrongly coupled to quantized modes is entangled and contains an arbitrary number of virtual photons. The problem of their detection has been raised since the very birth of the field, but despite the theoretical efforts still awaits experimental demonstration. Recently experimental problems have been addressed in detail showing that they can be overcome by combining an unconventional design of the artificial atom with advanced coherent control. In this work we study a simple scheme of control-integrated continuous measurement which makes remarkably favourable the tradeoff between measurement efficiency and backaction showing that the unambiguous detection of virtual photons can be achieved within state-of-the art quantum technologies.
} 
\maketitle

\section{Introduction}
\label{intro}
Advancement of fabrication technologies has allowed producing solid-state systems which exhibit the physics of atoms ultrastrongly coupled (USC) to quantized modes~\cite{ka:05-ciuti-prb-intersubbandpolariton,kr:19-forndiazsolano-rmp-usc,kr:19-kokchumnori-natrevphys-usc,20-leboite-aqt-theomethUSC}.
In these systems, the coupling constant $g$ between the artificial atom (AA) and the mode is comparable or even larger than the bare excitation frequencies $\epsilon$ and $\omega_c$ of the AA and of the mode.
This regime has been achieved on several different platforms~\cite{kr:19-forndiazsolano-rmp-usc,kr:19-kokchumnori-natrevphys-usc}, semiconductors~\cite{ka:09-anapparabeltram-prb-ustr,ka:09-guentertredicucci-nature-switcusc,ka:12-scalari-science-USCTHz} and superconductors~\cite{ka:10-niemczyck-natphys-ultrastrong,ka:10-diazmooji-prl-blochsiegert,ka:18-magazzu-natcomms-spinboson} being the most promising for applications.

In the USC regime, nonperturbative physics is predicted to emerge~\cite{kr:19-forndiazsolano-rmp-usc,kr:19-kokchumnori-natrevphys-usc} which is undetectable in the standard strong-coupling regime of quantum optics~\cite{kb:06-harocheraimond,ka:04-wallraff-superqubit}.
Higher-order antiresonant terms in the Hamiltonian break the conservation of the number of excitations this occurrence being at the heart of most of the phenomenology USC is expected to exhibit.
A striking feature is that eigenstates are highly entangled, and in particular the ground-state $\ket{\Phi}$ contains virtual photons (VP). The simplest instance is the two-level quantum Rabi model~\cite{ka:137-rabi-pr-rabimodel,ka:11-braak-prl-rabisol}
\begin{equation}
  \label{eq:rabiH}
  H_{R} = \epsilon \,\ketbra{e}{e} + g \,(a^\dagger + a) \, \left(\ketbra{g}{e} + \ketbra{e}{g}\right) + \omega_c \, a^\dagger a
\end{equation}
where $\{\ket{g},\ket{e}\}$ are the eigenstates of a two-level atom and $a$ ($a^\dagger$) are the annihilation (creation) operators of a quantized harmonic mode. The ground state has the form
\begin{equation}
  \ket{\Phi} = \sum_{n=0}^\infty \left(\ket{2n \,g} \braket{2n \,g}{\Phi}+ \ket{2n+1 \,e} \braket{2n+1 \,e}{\Phi}\right)
  \label{eq:rabi-gs}
\end{equation}
$\{\ket{n}\}$ being the number eigenstates of the mode \cite{kr:19-forndiazsolano-rmp-usc,kr:19-kokchumnori-natrevphys-usc}.
It is seen that $\ket{\Phi}$ contains an even number $\hat N = \hat n + \ketbra{e}{e}$ of excitations while in the absence of USC the ground state is $\ket{0g}$ thus it is factorized and it does not contain VPs.
The question of how to detect ground-state VPs in USC systems has been posed since the birth of the field~\cite{ka:05-ciuti-prb-intersubbandpolariton}.  They cannot be probed by standard photodetection because the USC vacuum $\ket{\Phi}$ cannot radiate~\cite{ka:18-distefanonori-scirep-qphotodetection}, thus
VPs must be converted to real excitations which are then detectable.
Early theoretical proposals of VP detection leverage time-dependent coupling constants as in the dynamical Casimir effect~\cite{ka:06-ciuticarusotto-pra-inputoutput,ka:07-deliberato-prl-uscdynamics,ka:09-deliberato-pra-extracavity} but require modulation of quantum hardware at subnanosecond times which is still unavailable.
Another class of proposals formulated in the last decade~\cite{ka:13-stassisavasta-prl-USCSEP,ka:14-huanglaw-pra-uscraman,ka:17-falci-fortphys-fqmt,ka:17-distefanosavastanori-njp-stimemission,ka:19-falci-scirep-usc} introduces an additional lower energy AA level $\ket{u}$ not coupled to the mode (see Fig.~\ref{fig:1}a) making $\ket{\Phi}$ a false vacuum which can undergo radiative decay.
This option also poses several experimental challenges, from the low yield of detectable photons to the fact that conventional quantum hardware does not ensure that the conversion is faithful, i.e. that output photons are really produced due solely to USC~\cite{ka:19-falci-scirep-usc}. For these reasons, despite the huge theoretical effort, detection of VPs
still awaits demonstration. In a recent work~\cite{ka:giannelli-vpdesign}, it has been shown that the above experimental problems can be overcome. Efficient, faithful and selective conversion of VPs to real ones can be achieved by combining an unconventional superconducting multilevel AA design~\cite{ka:19-hazard-prl-superinductances,ka:20-peruzzofink-prapplied-superinductance,ka:19-zhanggershenson-prappl-superinductor}, with coherent amplification of the conversion of ground-state VPs by advanced control~\cite{ka:17-falci-fortphys-fqmt} and a tailored measurement protocol.

In this work, we discuss a toy model of integrated protocol combining VP conversion by STIRAP~\cite{ka:giannelli-vpdesign} with photodetection by continuous measurement of the mode~\cite{ka:18-distefanopaternostro-prb-measurethermo}. In particular, we consider decay into a transmission line coupled to the mode during the whole protocol, which is the simplest experimental option. A detectable signal is obtained if the decay rate $\kappa$ of the mode is large enough which however, determines a backaction of the continuous measurement inducing decoherence which may affect the efficiency of the coherently amplified conversion. Our results show that STIRAP is resilient to this backaction making favourable the tradeoff of the integrated protocol with a continuous measurement.

\section{Model}
\label{sec:1}
We illustrate VPs conversion/detection considering the Hamiltonian of a three-level AA  coupled to the mode
(see Fig.~\ref{fig:1}a)
\begin{equation}
  \label{eq:H-3level}
  H = H_R -\epsilon^\prime \,\mathbbm{1}_{osc} \otimes \ketbra{u}{u} +  \omega_c \,a^\dagger a \otimes \ketbra{u}{u}.
\end{equation}
Hamiltonian~\eqref{eq:H-3level} describes a three-level AA with the two excited states $\ket{g}$ and $\ket{e}$ ultrastrongly coupled to the mode, as described in Eq.~\eqref{eq:rabiH}, and the ground state $\ket{u}$ uncoupled. It is a three-level approximation of the multilevel Hamiltonian described in ref.~\cite{ka:giannelli-vpdesign}, which is implemented by a fluxonium-like superconducting AA galvanically coupled to a mode also implemented by a superconducting LC resonator. Parameters for the lumped elements of this quantum circuit have been found such that coupling of $\ket{u}$ to the mode is very small thus guaranteeing that only VPs are converted in real photons. For instance, the first level splitting $\epsilon^\prime > 0$ of the \textquote{uncoupled} state $\ket{u}$ is much larger than $\omega_c = \epsilon$, the latter being the second atomic splitting. For a detailed discussion we refer to ref.~\cite{ka:giannelli-vpdesign}.
The eigenstates of $H$ are classified in two sets (see Fig.\ref{fig:1}a), namely the factorized states $\{\ket{n} \otimes \ket{u}\}$ with energies $-\epsilon^\prime + n \omega_c$,  and the entangled Rabi-like eigenstates $\{\ket{\Phi_l}\}$ of $H_R$, with eigenvalues $E_l$.

\subsection{Conversion of VPs}
\label{sec:conversion}
\begin{figure}[t]
  \begin{minipage}{0.40\columnwidth}
    (a)
  \end{minipage}
  \begin{minipage}{0.579\columnwidth}
    (b)
  \end{minipage}\\
  \begin{minipage}{0.40\columnwidth}
    \includegraphics[width=1\textwidth]{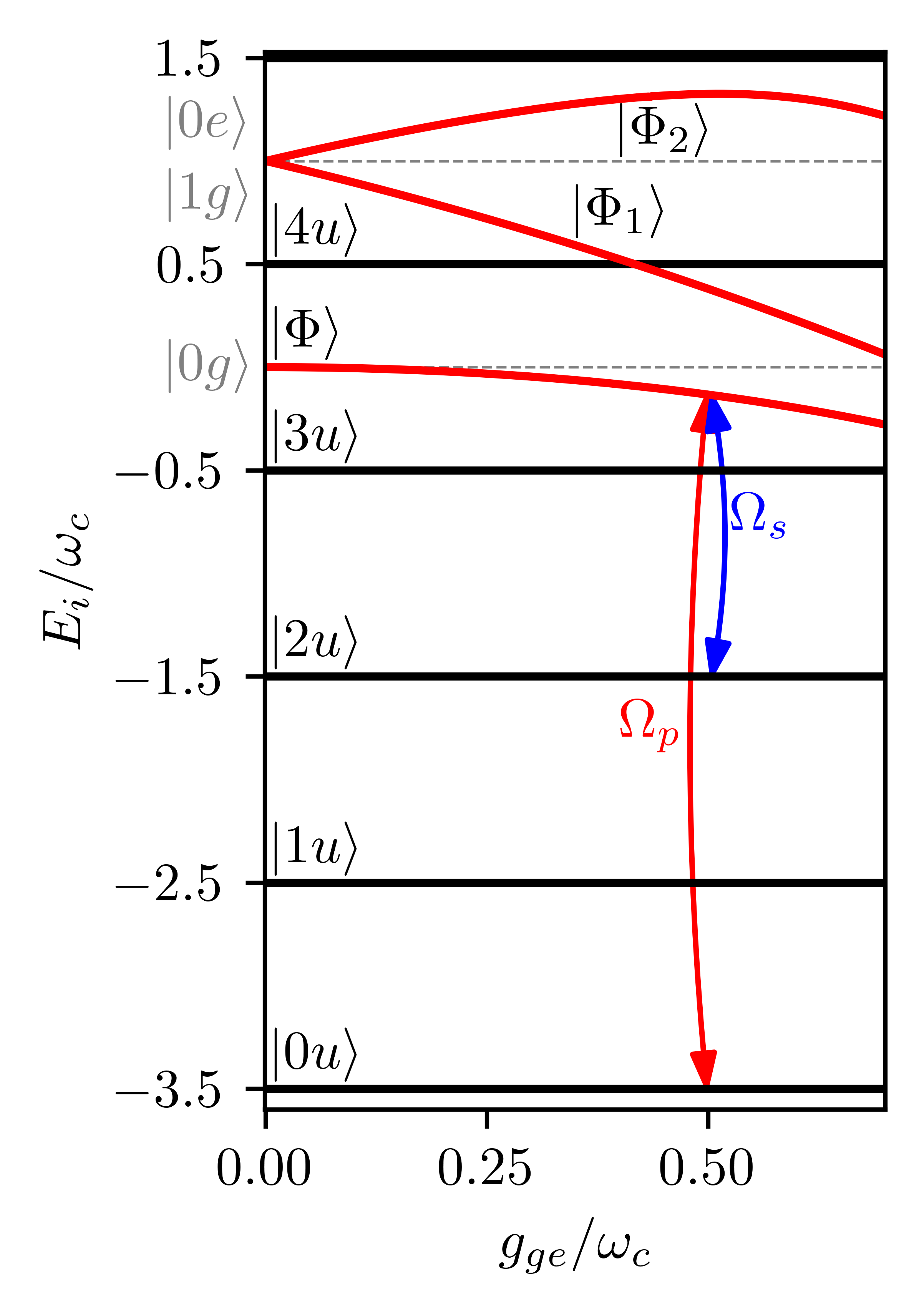} 
  \end{minipage}
  \begin{minipage}{0.53\columnwidth}
    \includegraphics[width=0.99\textwidth]{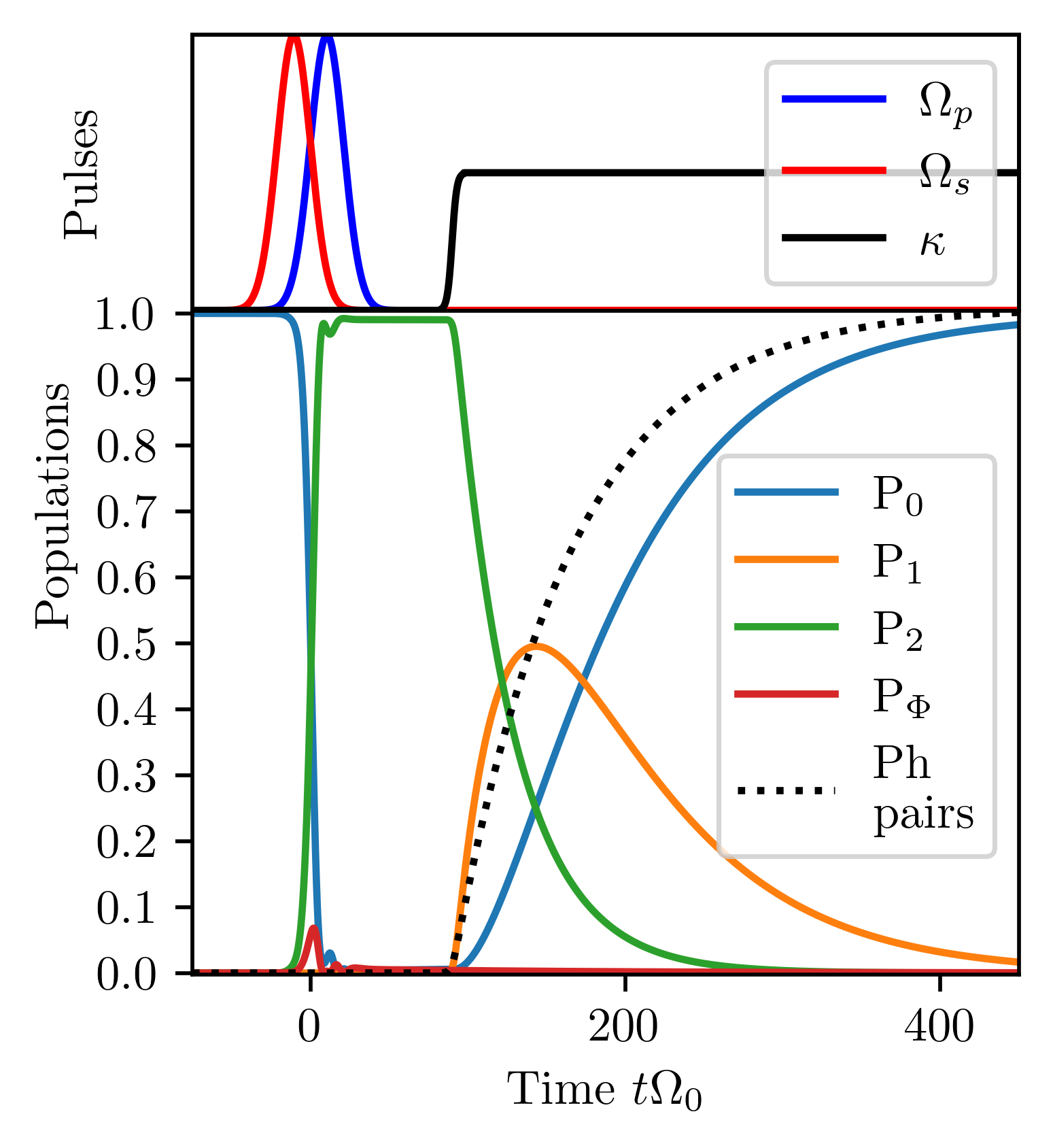} 
  \end{minipage}
  \caption{(a) Spectrum of the multilevel AA extended Rabi model with the additional uncoupled state $\ket{u}$, Eq.(\ref{eq:H-3level}), as a function of $g$ and the scheme of the Lambda configuration used to drive the system, Eq.(\ref{eq:control-lambda}). (b) Top panel: Gaussian pulses $\Omega_{s/p}(t)$ in the counterintuitive sequence for the coherent amplification protocol and the sigmoid function $\kappa(t-t_{sm})$ mimicking a switchable meter.
    Bottom panel: population histories; for $t<t_{sm}$ population transfer $\ket{0u} \to \ket{2u}$ by STIRAP is completed;  for $t>t_{sm}$ the mode decays, emitting two photons and resetting the system to the initial state. Parameters are given in  Table \ref{tab:parameters}. The protocol starts
    at $\Omega_0 t_{i}=-75$ and ends at  $\Omega_0 t_f=450$, the meter is switched on at $\Omega_0 t_{sm}=90 $. We used the shorthand notation for the populations $P_n:=\braket{n,u|\rho(t)}{n,u}$ for $n=0,1,2$ and $P_\Phi:=\braket{\Phi|\rho(t)}{\Phi}$, where $\rho(t)$ is the density matrix of the system.}
  \label{fig:1}       
\end{figure}

Conversion of VPs employs a STIRAP protocol~\cite{kr:17-vitanovbergmann-rmp} where the system is driven by a two-tone field
$W(t)=\mathscr{W}_s(t) \cos \omega_s t + \mathscr{W}_p(t) \cos \omega_p t$
mainly coupled with the $u-g$ transition of the AA. We take the field resonant with the two relevant transitions $\omega_p \approx E_0 + \epsilon^\prime$ and $\omega_s \approx E_0 + \epsilon^\prime - 2 \omega_c$. Standard approximations yield the $\Lambda$ driving  configuration~\cite{kr:01-vitanov-advatmolopt} of
Fig.~\ref{fig:1}a described in a rotating frame by the control Hamiltonian~\cite{ka:17-falci-fortphys-fqmt}
\begin{equation}
  \label{eq:control-lambda}
  \tilde{H}_C(t) = \frac{\Omega_p(t)}{2}
  \, \ketbra{0u}{\Phi} +
  \frac{\Omega_s(t)}{2}
  \, \ketbra{2u}{\Phi}+ \mbox{h.c.},
\end{equation}
where the Rabi amplitudes
$\Omega_p(t) =
  \mathscr{W}_p(t) \, \gamma_{ug}\, \braket{0g}{\Phi}
$ and
$\Omega_s(t) =
  \mathscr{W}_s(t) \, \gamma_{ug}\, \braket{2g}{\Phi}$
depend on the matrix element $\gamma_{ug} := \braket{u|\hat \gamma}{g}$ of the AA ``dipole'' operator~\cite{ka:giannelli-vpdesign}. Operating the ``counterintuitive'' pulse sequence~\cite{kr:17-vitanovbergmann-rmp} $\Omega_{p/s}(t)= F[(t\mp \tau)/T_W]$ with $\tau>0$, the Stokes pulse is shined before the pump pulse (see Fig.\ref{fig:1}b, top panel).
Using for instance Gaussian pulses of width $T_W$, coherent population transfer $\ket{0u} \to \ket{2u}$ occurs with $\sim 100\%$ probability provided the ``global adiabaticity'' condition $\max_t[\Omega_s(t)] T_W \gtrsim 10$~\cite{ka:22-giannelli-pla-tutorial,kr:98-bergmann-rmp-stirap} is met~\cite{ka:17-falci-fortphys-fqmt}.
Population transfer may occur only if $\Omega_s(t) \propto  \braket{2g}{\Phi} \neq 0$ thus it provides a ``smoking gun'' of the presence of VPs in the ground-state. In the target state $\ket{2 u}$, two real photons are present, witnessing the presence of the two-VPs component in $\ket{\Phi}$. Therefore, this protocol guarantees 100\% conversion efficiency thanks to coherence. In this case, the dynamics is restricted to the subspace spanned by the eigenstates $\{\ket{0u},\ket{2u},\ket{\Phi}\}$.
Population histories are shown in the lower panel of Fig.~\ref{fig:1}b, population transfer by STIRAP occurring in the first part of the protocol.
As shown in Ref.~\cite{ka:giannelli-vpdesign} the nearly ideal scenario described in this section can be implemented also in the USC regime by state-of-the-art superconducting quantum technologies in an unconventional design of the quantum circuit, and using superinductors~\cite{ka:19-hazard-prl-superinductances,ka:20-peruzzofink-prapplied-superinductance,ka:19-zhanggershenson-prappl-superinductor} and advanced control at microwave frequencies. Control based on STIRAP has been proposed~\cite{ka:13-falci-prb-stirapcpb,ka:15-distefano-prb-cstirap,ka:16-distefano-pra-twoplusone,ka:16-vepsalainen-photonics-squtrit} and demonstrated~\cite{ka:16-kumarparaoanu-natcomm-stirap,ka:16-xuhanzhao-natcomm-ladderstirap} in standard superconducting quantum devices.

\subsection{Toy-model for photodetection}
\label{sec:2}
Ideally, once the population has been transferred in $\ket{2u}$ the converted VP pair can be detected.
For oscillators with quality factor $Q \gtrsim 10^4$ the population of the mode remains large enough~\cite{ka:21-ridolfo-epjst-virtualphotonopen} to allow photons to be detected (and even counted) by single-shot nondemolition measurements performed by a quantum probe coupled dispersively to the mode~\cite{ka:11-eichlerwallraff-prl-photontomography,ka:21-curtisschoelkopf-pra-pnrdetection}. A much simpler procedure is a continuous measurement~\cite{ka:18-distefanopaternostro-prb-measurethermo} which uses radiative decay of converted VPs with a rate $\kappa$ into a transmission line. In this case, a key advantage is that
the initial state is faithfully prepared by simply letting the system relax~\cite{ka:12-spagnolo-aphyspol-relaxation} due to photodetection. Thus the protocol can be repeated over and over yielding a detectable signal if $\kappa$ is large enough.

Since the oscillator selection rules prevent direct $\ket{2u} \to \ket{0u}$ decay, photodetection involves the sequential decay of the mode $\ket{2u} \to \ket{1u} \to \ket{0u}$.
Therefore we formulate a minimal model restricting the analysis to the four-dimensional Hilbert space spanned by $\left\{\ket{0u},\, \ket{1u},\,\ket{2u},\, \ket{\Phi}\right\}$.
In Fig.~\ref{fig:1}b we show the population histories with
an oscillator decay rate $\kappa(t)$ turned on after the completion of STIRAP. Photons in $\ket{2u}$ first decay to $\ket{1u}$ and then to $\ket{0u}$ at each step a photon being emitted into the transmission line.
A minimal model of decay is described by
a Lindblad equation with Lindbladian given by
\begin{equation} \label{eq:lindbladian-cavity}
  {\cal L}_\kappa = \kappa \, {\cal D}[\hat a] + \kappa \, \mathrm{e}^{- \beta \omega_c}\, {\cal D}[\hat a^\dagger]
\end{equation}
where the dissipator is defined as ${\cal D}[\hat A] \hat{\rho} = \hat A \hat \rho \hat A^\dagger - \frac{1}{2} \big(\hat A^\dagger \hat A\, \hat \rho + \hat \rho \,\hat  A^\dagger\hat  A \big)$ and $\hat{\rho}$ is the density matrix of the system. The first term describes emission with decay rate $\kappa$ and the second describes absorption whose rate has been written making the phenomenological assumption that detailed balance at thermal equilibrium, $\beta =1/(k_B T)$, can be used also for the driven system.

We point out that some care is required to guarantee a physically consistent picture of photon decay.
We now briefly describe how to interpret Eq.(\ref{eq:lindbladian-cavity}) to obtain the minimal model of photodetection.
The key point is that the operator $\hat a$ must be defined such to avoid photon annihilation in the Rabi ground state $\ket{\Phi}$ otherwise we could have photon emission even in the absence of the level $\ket{u}$, which is unphysical. The complete theory requires using ``dressed'' field operators, say $\hat a \to \hat{\mathrm{A}}$ \cite{kr:19-kokchumnori-natrevphys-usc}. In our case, this brings a simplification since the new operators are such that  $\hat{\mathrm{A}}\ket{\Phi}=0$ and they reduce to $\hat a$ when acting on $\ket{nu}$. Since truncation to the four-level space also implies that $\hat{\mathrm{A}}^\dagger \ket{\Phi} =0$ we simply have to use a projected version of the bare jump operators acting only on states $\ket{nu}$. This provides the correct minimal description of both decoherence and photodetection.

The measured quantity is the extra current of photons emitted into the transmission line defined as
\begin{equation}
  \label{eq:total-phcurrent}
  j_T = \kappa \, \big[\overline{\langle \hat a^\dagger(t) \hat  a(t)\rangle} - \langle \hat a^\dagger \hat a\rangle_{th} \big]
\end{equation}
where the first term is the average total current for the driven system
$$
  \overline{\langle \hat a^\dagger(t)  \hat a(t)\rangle}
  = \frac{1}{ t_M} \int_0^{t_M} \hskip-3pt dt
  \langle \hat a^\dagger(t)\,  \hat a(t)\rangle
$$
where $t_M$ is the duration of the whole integrated conversion and measurement protocol and the second term is the equilibrium thermal current in the undriven system. Fig.~\ref{fig:1}b also shows the number of emitted photon pairs (black dashed line) in a cycle at zero temperature when the thermal current is also zero. The parameters used are such that the VPs conversion is complete and the system is reset to the initial state $\ket{0u}$.
\begin{table}
  \label{tab:1}       
  \centering
  \begin{tabular}{cccccccc}
    \hline\noalign{\smallskip}
    $\Omega_{p/s}^\mathrm{max}$ & $T_W$                 & $\tau$                  & $1/\kappa$ & $\gamma$ & $\epsilon$            & $\epsilon^\prime$       & $T$                               \\
    \noalign{\smallskip}\hline\noalign{\smallskip}
    $\Omega_0$                  & $\frac{15}{\Omega_0}$ & $\frac{10.5}{\Omega_0}$ & $5T_W$     & $\kappa$ & $\hbar \omega_c$      & $5.9 \, \hbar \omega_c$ & $\frac{\hbar \omega_c}{1.95 k_B}$ \\
    \noalign{\smallskip}\hline\noalign{\smallskip}
    $50 \times 2 \pi\,$         & 48                    & 33.6                    & 240        &          & $2.03 \times 2 \pi\,$ & $11.98 \times 2 \pi\,$  & $50\,$                            \\
    MHz                         & ns                    & ns                      & ns         &          & GHz                   & GHz                     & mK                                \\
    \noalign{\smallskip}\hline
  \end{tabular}
  \caption{ Parameters used in the simulations and corresponding physical values for the superconducting flux-based architecture including a superconducting junction with Josephson energy $E_J/(2 \pi)= 10 \,$GHz considered in Ref.\cite{ka:giannelli-vpdesign}, where an extended Rabi model was studied with coupling constant $g/\omega_c =0.5$ yielding squared matrix elements $|\braket{0g}{\Phi}|^2=0.42$ and $|\braket{2g}{\Phi}|^2= 0.05$ for the probability of finding respectively $n=0,2$ virtual photons in the ground state.}
  \label{tab:parameters}
\end{table}

\subsubsection{Atomic decay}
In principle, atomic decay is not relevant in the ideal protocol since in STIRAP the intermediate state $\ket{\Phi}$ is expected not to be populated (see Fig.~\ref{fig:1}b). However, we will see that AA decay  plays a role in the integrated conversion/measurement protocol (see Fig.~\ref{fig:2}a). Thus we take into account it in the Lindblad formalism by introducing two dissipators with jump operators $\ketbra{0,u}{\Phi}$ and $\ketbra{2u}{\Phi}$.
Nonradiative decay rates could be explicitly calculated by the Fermi Golden Rule if the atomic environment were specified. Even in the absence of detailed information, we know that the rates $\ket{\Phi} \to \ket{n u}$ are proportional to the square of the matrix elements
$
  \bra{nu} \big[ \mathbbm{1}_{osc} \otimes \ketbra{u}{g} \big] \ket{\Phi} =
  \braket{n g}{\Phi}
$, thus once again we express them in terms of a single parameter $\gamma$ as
$$\gamma_{0\Phi} = |\braket{0 g}{\Phi}|^2\,\gamma \qquad ;\qquad  \gamma_{2\Phi}= |\braket{2 g}{\Phi}|^2 \, \gamma \;.$$
The excitation rates at equilibrium are written assuming again that detailed balance holds and finally, we obtain the Lindbladian
\begin{multline}\label{eq:lindbladian-AA}
  {\cal L}_\gamma = \gamma_{0\Phi} \,{\cal D}\big[\ketbra{0u}{\Phi}\big]
  + \gamma_{0\Phi} \,\mathrm{e}^{- \beta \epsilon^\prime} \, {\cal D}\big[\ketbra{\Phi}{0u}\big]\\
  + \gamma_{2\Phi} \,{\cal D}\big[\ketbra{2u}{\Phi}\big]
  + \gamma_{2\Phi} \,\mathrm{e}^{- \beta (\epsilon^\prime-2 \omega_c)}\,{\cal D}\big[\ketbra{\Phi}{2u}\big].
\end{multline}

\section{Results and conclusions}
\label{sec:results}
\begin{figure}[t]
  \centering
  \begin{minipage}{0.60\columnwidth}\hspace{30pt}(a)\\
    \includegraphics[width=\textwidth]{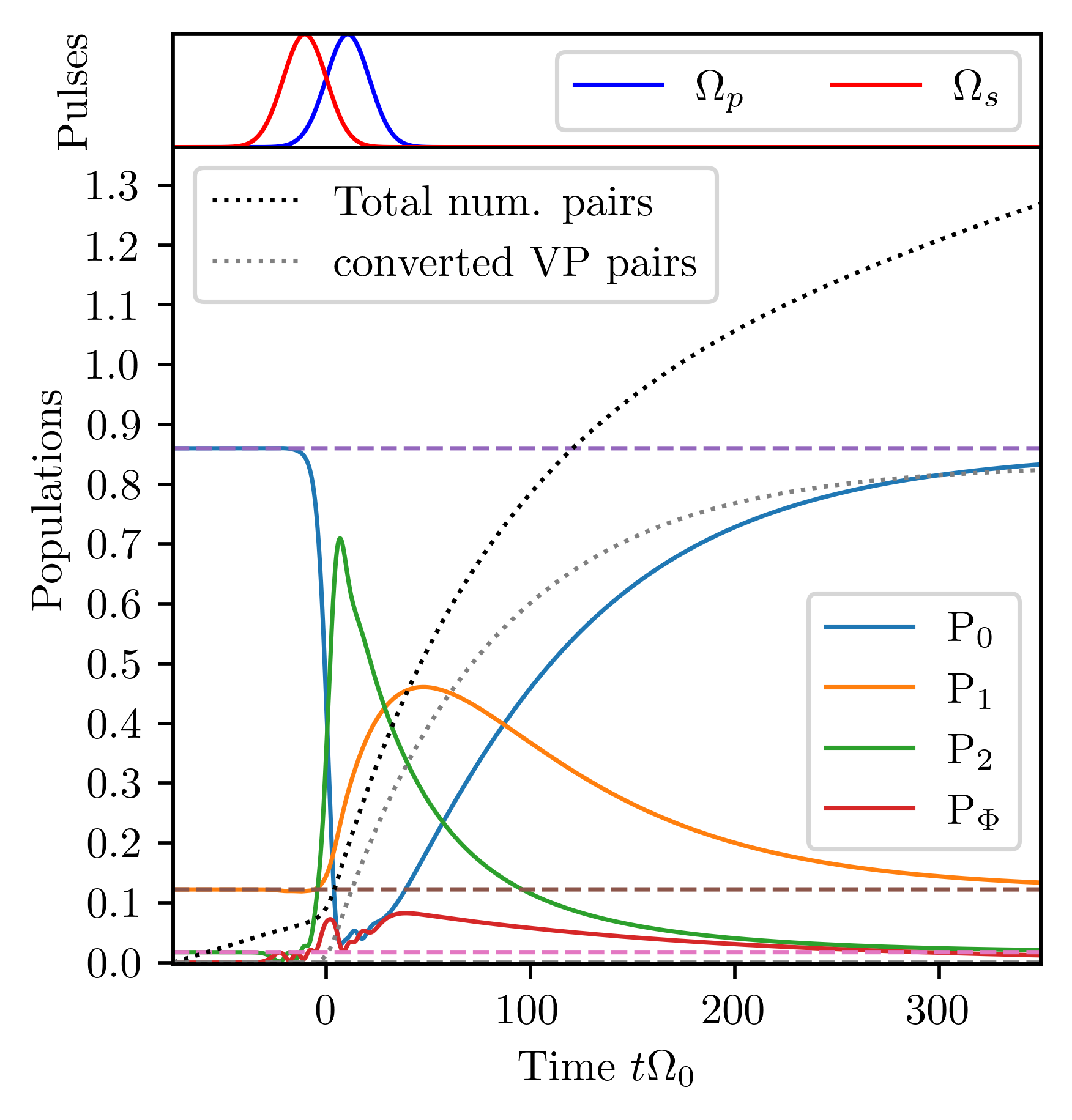}
  \end{minipage}
  \begin{minipage}{0.37\columnwidth}\hspace{40pt}(b)\\
    \includegraphics[width=\textwidth]{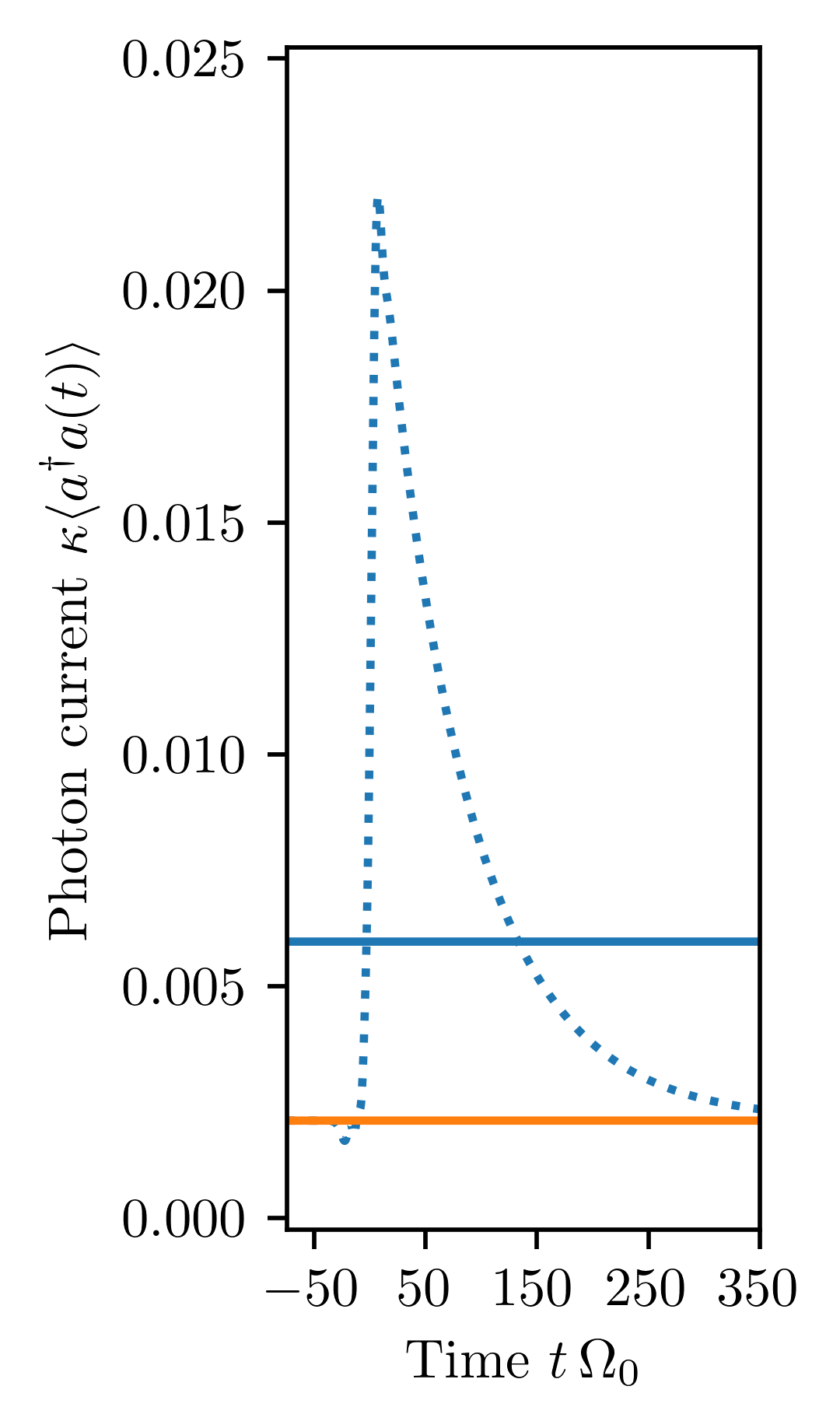}
  \end{minipage}
  \caption{
    (a) A cycle of integrated conversion/photodetection protocol at finite temperatures. The meter is always on, i.e. $\kappa \neq 0$ (not represented) is independent on time, allowing for an overall time $t_M$ shorter than in
    Fig.\ref{fig:1} (here $\Omega_0 t_f=350$). Parameters are given in  Table \ref{tab:parameters} and for the case study considered $t_M= 425/\Omega_0=1.36\,\mu$s.
    The black dotted line is the total number of photons decaying in the transmission line: it becomes larger than two because of the thermal contribution which is linear in $t$.
    By subtracting it, we obtain the number of converted VPs (grey dotted line). The four horizontal dashed lines are the thermal populations.
    (b) The instantaneous emitted current (dotted line) and its average per cycle (blue solid) compared to the thermal current (orange).}
  \label{fig:2}       
\end{figure}

The main result of this work is shown in Fig.\ref{fig:2}a where we plot population histories at finite $T=50\,$mK for an always-on detector, i.e. taking $\kappa$ constant in our equations. Having in mind repeating the cycle over and over in order to collect a large enough signal for detection, we show a cycle starting from the thermal state of the system (horizontal dashed lines) that for the parameters chosen also ends in the same state, being thus a limiting cycle. Thermal effects reduce the net population transferred, thus the system converts pairs of VPs with a smaller probability. The backaction of the always-on detector is expected to further reduce population transfer and VP conversion due to dephasing during adiabatic passage in STIRAP and to decay of the mode when $\ket{2u}$ starts to be populated. The former effect leads to a reduction of the final population of $\ket{2u}$ after the completion of STIRAP estimated by   $P_{2} = \frac{1 }{ 3} + \frac{2 }{3} \mathrm{exp} \big[- 3 \kappa_\phi T^2/(16 \tau)]$~\cite{ka:04-ivanov-pra-stirapdephasing} where in our case $\kappa_\phi = 3 \kappa/2$~\cite{ka:giannelli-vpdesign}, which for the value of
$\kappa$ in Table\ref{tab:parameters} turns out to be small. On the contrary, decay of the mode after the adiabatic passage phases has a significant impact on ideal STIRAP since it determines in a strong reduction of $P_2$ (see
Fig.~\ref{fig:2}a). However, this population loss results in the detection of converted VPs photon pairs which progressively populate $\ket{2u}$ during STIRAP thus the probability of detecting a photon pair per cycle remains large. Notice that
the total number of photons decaying in the transmission line
(black dotted line in Fig.\ref{fig:2}a) is larger than two and increases linearly at very short and large times. This linear component is due to the constant current of thermal photons $\kappa \langle a^\dagger a \rangle_{th}$ which has nothing to do with VPs. By subtracting the thermal part, we obtain the number of detected VP pairs, which turns out to be almost equal to the thermal population of the initial state $\ket{0 u}$ (gray dotted line in Fig.\ref{fig:2}a).

Comparing the continuous measurement of Fig.\ref{fig:2}a with the switchable probe protocol of Fig.\ref{fig:1}b we notice that photodetection during the protocol strongly modifies population histories. However, coherent amplification of the VP is preserved since the population of $\ket{2 u}$ decaying before the completion of STIRAP is also due to converted VPs which are being detected. Besides being much simpler to implement, the continuous measurement scheme is faster
(notice the time scales in Fig.\ref{fig:2}a and Fig.\ref{fig:1}b) lowering the relative contribution of the stray thermal current. This is apparent in Fig.\ref{fig:2}b where we plot for each cycle after thermalization the instantaneous photon current (blue dotted) and the averaged total (blue) and thermal (orange) currents which are proportional to the power respectively emitted into the transmission line. Their difference is the signal due to the converted VPs, which is related to the grey dotted line in Fig.\ref{fig:2}a. Summing up, for the continuous measurement scheme the tradeoff between efficient measurement and decoherence is positive, yielding a sufficiently large extra output power from converted VPs.

We briefly comment on the role of atomic decay. Fig.~\ref{fig:2}a shows that in the integrated protocol, some population appears in $\ket{\Phi}$ before the completion of STIRAP since the pump pulse repumps to $\ket{\Phi}$ population which just decayed in $\ket{0u}$ because of the always-on coupling to the meter. In the absence of atomic decay, this population would remain trapped in $\ket{\Phi}$ after the completion of STIRAP. If $\gamma \neq 0$ this population relaxes non-radiatively to $\ket{0u}$ resetting efficiently the system for the next cycle.

In Ref.~\cite{ka:giannelli-vpdesign} it has been estimated that a signal corresponding to the case study of Table~\ref{tab:parameters}a could be amplified by standard HEMT circuitry and discriminated from thermal noise, this task requiring hundreds of repetitions of the conversion/measurement cycle, which is a reasonable figure. In this work, we have analysed in detail the dynamics of the continuous measurement, showing that STIRAP is resilient to measurement backaction.
It is likely that combining optimal control theory and advanced computational methods of data analysis~\cite{ka:21-brown-njp-rlcoherentpop,ka:22-giannelli-pla-tutorial} yields even better figures. For instance, Fig.~\ref{fig:2}b suggests that shortening the after-STIRAP part of the protocol yields a larger average output power for the converted VPs, while the thermal floor is unchanged. However, the steady-state population of $\ket{0 u}$ may be smaller and that population of the intermediate state may trigger leakage from the four-level subspace, requiring an analysis which takes into account the multilevel nature of the setup proposed in Ref.~\cite{ka:giannelli-vpdesign} and the subtleties of the physics of an open system in the USC regime.

\subsubsection*{Acknowledgements}
We acknowledge J. Rajendran and A. Ridolfo who helped to develop our insight for this paper.
\\
This work was supported by the QuantERA grant SiUCs (Grant No. 731473), by the
PNRR MUR project PE0000023-NQSTI, by ICSC–Centro Nazionale di Ricerca in
High-Performance Computing, Big Data and Quantum Computing, by the University of
Catania, Piano Incentivi Ricerca di Ateneo 2020-22, project Q-ICT. EP
acknowledges the COST Action CA 21144 superqumap and GSP acknowledges financial
support from the Academy of Finland under the Finnish Center of Excellence in
Quantum Technology QTF (projects 312296, 336810, 352927, 352925).

\subsubsection*{Data availability statement}
The datasets generated during and/or analysed during the current study are available from the corresponding author on reasonable request.
\printbibliography

\end{document}